\documentclass[preprint,review,12pt]{elsarticle}

\usepackage{amssymb}
\usepackage{amsmath}
\usepackage{lipsum}

\journal{Journal of Quantitative Spectroscopy and Radiative Transfer}

\begin{document}

\begin{frontmatter}



\title{Frequency redistribution and step-size distribution of light scattered by atomic vapor: applications to Lévy flight random walk.}


\author[1]{I. C. Nunes}
\author[2]{M. O. Araújo}
\author[1]{J. P. Lopez}
\author[1]{T. Passerat de Silans}

\affiliation[1]{organization={Departamento de Física, CCEN},
addressline={Universidade Federal da Paraíba, Caixa Postal 5008},
postcode={58051-900},
city={João Pessoa, Paraíba},
country={Brazil}}
\affiliation[2]{organization={Departamento de Física},
addressline={Universidade Federal de Pernambuco},
postcode={50670-901},
city={Recife, Pernambuco},
country={Brazil}}

\begin{abstract}
The propagation of light that undergoes multiple-scattering by resonant atomic vapor can be described as a Lévy flight. Lévy flight is a random walk with heavy tailed step-size (r) distribution, decaying asymptotically as $P(r)\sim r^{-1-\alpha}$, with $\alpha<2$. The large steps, typical of Lévy flights, have its origins in frequency redistribution of the light scattered by the vapor. We calculate the frequency redistribution function and the step-size distribution for light diffusion in atomic vapor. From the step-size distribution we extract a Lévy parameter $\alpha$ that depends on the step's size. We investigate how the frequency redistribution function and step-size distribution are influenced by the finite size of the vapor and the many-level structure typical for alkali vapors. Finite size of the vapor introduces cutoff on the light scattered spectrum and thus in the size of steps. Multi-level structure introduces oscillations in $P(r)$ slope. Both effects might have an impact on measurables related to the Lévy flight random walk.
\end{abstract}



\end{frontmatter}

\section{Introduction}

Random walks are present in a large variety of systems, with the Brownian motion of particles on water surfaces \cite{Barnes1934,Bouchaud1990} the most known example. Brownian motion is characterized by a random walk with successive jumps of size $r$, and performed within a given distribution with finite step-size variance. In the last decades, random walks with diverging step-size variance, known as Lévy flights or Lévy walks, have been extensively studied. Lévy flights are random walks characterized by a non-negligible probability of occurring a large step. The step-size probability distribution decays asymptotically as $P(r)\propto r^{-1-\alpha}$, with the Lévy parameter $\alpha<2$, while for normal random walk we have $\alpha>2$.\\

In Lévy flights, the particles perform instantaneous jumps (infinity velocity) while for Lévy walks the jump sizes and their duration are connected by the particle speed (usually considered as constant) \cite{Zaburdaev2015}. Many systems have been shown to perform Lévy flights or Lévy walks, e.g., animal foraging \cite{Viswanathan1996}, phonons transport in nanowires \cite{Li2022} and spread of genes lineage \cite{Smith2023}. Another example of Lévy flight is light scattering in resonant atomic vapor. The random walk can be described as a sequence of photon absorption by atoms in the vapor followed by spontaneous emission~\cite{Molisch1998}. The probability rest time distribution of the excitation of the atom is $P(t)\propto e^{-t/\tau}$, with the excited level lifetime $\tau$ of the order of tens of nanoseconds. Typical duration of steps for systems of size of $\sim 1$ m is ten time less. The duration of the steps are negligible and those random walks can be treated as Lévy flights \cite{Zaburdaev2015}.  

The existence of large steps in light scattering in atomic vapor has been recognized long ago as a contribution of photons emitted in the wings of the emission profile~\cite{Osterbrock1962,Adams1972}. The step-size distribution depends on the vapor emission and absorption profiles and was calculated for two-level atoms~\cite{Pereira2004} for Complete Frequency Redistribution (CFR). For CFR the emission and absorption profiles are the same $\Theta=\phi$, with $\Theta$ ($\phi$) the emission (absorption) profile. The calculated asymptotic values of Lévy $\alpha$ parameter were $\alpha=0.5$ for both Lorentz and Voigt profiles and $\alpha\approx1.0$ for Doppler profile \cite{Pereira2004} and these values were verified experimentally \cite{Mercadier2009,Mercadier2013,Baudouin2014,Araujo2021,Macedo2021}. Furthermore, a step size dependent $\alpha$ parameter can be obtained by defining $1+\alpha(r)=-\frac{dlog(P(r))}{dlog(r)}$ \cite{Lopez2023}. From measurements of transmission through a Cs vapor it was shown, that the $\alpha$ parameter depends on the system size with $\alpha\approx 1$ for small system and $\alpha=0.5$ for large system \cite{Macedo2021}. For small systems, photons emitted in the Doppler core of the Voigt profile can escape from the sample resulting in a Doppler-like $\alpha$ value, while for large systems only photons emitted at the Lorentz wings escapes corresponding to $\alpha=0.5$. The results seems consistent with the \textit{Single-Big-Jump} Principle \cite{Vezzani2019} standing the light escapes the vapor after a single jump with the size of the vapor \cite{Adams1972} in the sense that the size of the system rules the measured $\alpha$ value. 

Frequency is not totally redistributed in a single scattering of the photon by an atom. Partial correlation between absorbed and emitted frequency exists due to \cite{Pereira2007,Carvalho2015}: (i) velocity of the absorbing atom and (ii) coherence of the scattering in atomic rest frame. Emission profiles for Partial Frequency Redistribution (PFR) has been calculated since the decade of 50s \cite{Unno1952,Hummer1962,Omont1972,Domke1988} with particular interest in astrophysical system, as, for instance, scattering of Ly $\alpha$ lines in optically thick nebulae \cite{Unno1952} and Ca I line in solar spectrum \cite{Frisch1996}. 

Experimental work on resonant vapors is a good platform both for studying Lévy flights and frequency redistribution. Experiments on Lévy flight have been performed using alkali vapors (Cs and Rb)\cite{Mercadier2009,Mercadier2013,Baudouin2014,Araujo2021,Macedo2021,Lopez2023}.  Those system are rich for the study of Lévy flights and allows to investigate: i) consequence to measurable quantities of step-size dependent $\alpha$ parameter value \cite{Chevrollier2012,Lopez2023}; possibility of changing the (effective) system size by 4 orders of magnitude \cite{Baudouin2014,Araujo2021,Macedo2021,Lopez2023}, by changing the vapor density; intermittence between two distinct step-size distributions \cite{Pereira2007,Lopez2023}. From the frequency redistribution point of view analyzing transmission allows to have information on emission at the profiles wings \cite{Adams1972}. Although experiments were performed in alkali vapor, calculations of step size distribution available in literature were performed for infinite-size vapor of two-level atoms\cite{Pereira2004,Pereira2007,Chevrollier2012,Mercadier2013}. Here, we calculate the step-size distribution of light diffusing in two-level atomic vapor, as well as, in Cs and Rb vapors. In Sect. 2 we describe how to calculate step-size distribution for photon scattering in atomic vapor. In Sect. 3 we discuss influence of the finite size of the system on the step-size distribution for two level-atomic vapor. Than, in Sect. 4 we investigate the influence of both ground and excited hyperfine manyfold of Cs $D_2$ and $D_1$ and Rb $D_2$ lines in the step-size distribution. Finally, in Sect. 5 we summarize our results.

\section{Calculation of step length distribution}

Let consider a two-level atom whose resonant frequency is $\nu_0$. The step size distribution for a photon at a given  detuning $\delta=\nu-\nu_0$ is given by \cite{Holstein1947,Pereira2004,Chevrollier2012}:
\begin{equation}
    P_\delta(l)=N\sigma(\delta)e^{-N\sigma(\delta)l}, \label{P_Beer_Lambert}
\end{equation}
with $N$ the atomic vapor density and $\sigma(\delta)$ the light-atom absorption cross section at detuning $\delta$. Step-size distribution in Eq. \ref{P_Beer_Lambert} has a mean step-size of $\bar{l}=(N\sigma(\delta))^{-1}$, that depends on photon frequency, resulting in a spectral inhomogeneity of the step-size distribution \cite{Zaburdaev2015}. To calculate the actual step-size distribution one has to average Eq. \ref{P_Beer_Lambert} by the probability of finding a photon with detuning $\delta$: $\Theta(\delta)$ \cite{Holstein1947,Pereira2004,Chevrollier2012}:
\begin{equation}
    P(l)=\int d\delta \Theta(\delta)N\sigma(\delta)e^{-N\sigma(\delta)l}. \label{P_l}
\end{equation}

Step-size distribution of Eq. \ref{P_l} can be also expressed in terms of a normalized absorption profile $\phi(x)$ and dimensionless size $r$, as \cite{Pereira2007}:
\begin{equation}
    P(r)=\int dx \Theta(x)\phi(x)e^{-\phi(x)r}, \label{P_r}
\end{equation}
with $\phi(x)=N\sigma(x)\frac{l}{r}$ and $r=N\frac{\sigma(0)}{\phi(0)}l$. Here we also introduce a detuning normalized by the Doppler width of the transition $x=\delta/\Gamma_D$ \cite{Hummer1962}. \\

The probability of finding a photon with frequency $\Theta(x)$ corresponds to the scattered spectral profile. The photons scattered by the vapor are redistributed in frequency and one can define \cite{Hummer1962} a redistribution function, $R(x,x')$, that is the joint probability that an incident photon with detuning between $x$ and $x+dx$ be scattered with detuning between $x'$ and $x'+dx'$, with normalization $\int R(x,x')dx'=\phi(x)$\cite{Hummer1962}. $R(x,x')$ was calculated by D. Hummer \cite{Hummer1962}  for four cases, and we will discuss two of them in this article, the so called R$_{II}$ and R$_{III}$ cases.

The $R_{II}$ case considers a coherent scattering in the atomic rest frame, that is, the redistribution function in the atomic rest frame is $R^{(at)}=\delta(x^{at}-x'^{at})$, with $\delta$ denoting the Dirac delta function and $x^{at}$ and $x'^{at}$ are incident and scattered detunings in the atomic rest frame, respectively. This case occurs when excited state is broadened by radiation damping and no external agent perturbs the atomic dipole. The frequency is further redistributed in the laboratory frame by Doppler shift due to change of the photon direction in the scaterring. The resultant redistribution function is \cite{Hummer1962}:
\begin{equation}
    R_{II}(x,x')=\frac{1}{\pi^{3/2}}\int_{\frac{1}{2}\left|\bar{x}-\underbar{x}\right|}^{\infty}e^{-y^2}\left[\tan^{-1}\left(\frac{\underbar{x}+y}{\sigma}\right)-\tan^{-1}\left(\frac{\bar{x}-y}{\sigma}\right)\right]dy,
\end{equation}
with $\bar{x}=max\left(x,x'\right)$ and $\underbar{x}=min\left(x,x'\right)$.

The R$_{III}$ case considers natural damping and elastic collisions between atoms. The frequency is totally redistributed in the atomic rest frame with emission profile \cite{Hummer1962}:
\begin{equation}
    f(x'^{at})=\frac{2\Gamma}{\pi}\frac{1}{\left(x'^{at}\right)^2+4\Gamma^2}, \label{lorentzian}
\end{equation}
with $\Gamma=\Gamma_n+\Gamma_C$ the homogeneous broadening of the transition taking into account natural broadening $\Gamma_n$ and collisional broadening $\Gamma_C$. Note that, in the atomic rest frame, the detuning of the scattered photon is independent of incident detuning. Again, frequency is further redistributed in laboratory frame due to Doppler effect resulting in a redistribution function \cite{Hummer1962}: 

\begin{equation}
\begin{split}
    R_{III}(x,x')=\frac{1}{\pi^{5/2}}\int_{0}^{\infty}e^{-y^2}\left[\tan^{-1}\left(\frac{x+y}{\sigma}\right)-\tan^{-1}\left(\frac{x-y}{\sigma}\right)\right]\times\\
    \left[\tan^{-1}\left(\frac{x'+y}{\sigma}\right)-\tan^{-1}\left(\frac{x'-y}{\sigma}\right)\right]dy.
\end{split}
\end{equation}

\section{Two-level vapor with finite size}

A usual way of calculating the step-size distribution for light in resonant vapor is to consider CFR, which is reasonable if the photon undergoes many collisions during the scattering process \cite{Pereira2004,Chevrollier2012,Araujo2021, Macedo2021}. Partial Frequency redistribution was also considered \cite{Pereira2007,Mercadier2013,Lopez2023} by introducing a recurrence rule that gives the emission spectra at step $n$ from the emission spectra at step $n-1$ and from the redistribution function \cite{Pereira2007,Mercadier2013,Lopez2023}. To our knowledge, the recurrence rule was previously applied only for medium with infinite size \cite{Pereira2007,Mercadier2013,Lopez2023}. Here we consider a sample with size $L$ and see the influence of the system size in the step-size distribution.

In PFR, the scattered frequency depends on the incident frequency and scattered spectrum evolves with the number of scattering events. The conditional probability that a photon of frequency $x'$ is scattered given that a  photon of frequency $x$ is absorbed is given by  \cite{Hummer1962,Pereira2007,Mercadier2013,Lopez2023}:
\begin{equation}
    P(x'|x)=\frac{R(x,x')}{\phi(x)}.
\end{equation}

The normalized absorption profile $\phi(x)$ is given for a two-level vapor by a Voigt profile \cite{Pereira2007}:
\begin{equation}
\phi(x)=\frac{a}{\pi^{3/2}}\int_{-\infty}^{\infty}\frac{e^{-y^2}}{a^2+\left(x-y\right)^2}dy. \label{Voigt}    
\end{equation}

For calculating the scattered spectrum after scattering event of number $n$ one has to integrate $P(x'|x)$ over all possible incident frequency $x$ weighted by the probability that this frequency was emitted at step $n-1$ and by the probability that this photon was scattered by the vapor:
\begin{equation}
    \Theta(x')_n=\int dx \left[1-T(x)\right]\Theta_{n-1}(x)\frac{R(x,x')}{\phi(x)}. \label{Eq:recurrence}
\end{equation}

The factor $\left[1-T(x)\right]$ is introduced to take into account that only a fraction of the photons emitted in step $n-1$ will be scattered. We take $T(x)$ as the ballistic transmission of a photon through the sample of size $L$:
\begin{equation}
    T(x)=e^{-N\sigma(x)L}=e^{-\phi(x)r_L},
\end{equation}
with $r_L=N\frac{\sigma(0)}{\phi(0)}L$ the sample dimensionless size. The choice of $T(x)$ is justified from the \textit{Single-big-jump principle} \cite{Vezzani2019}: excitation remains close to the starting point of the random walk till a jump of size $r\sim r_L$ occurs. Thus the relevant size over which we have to verify if the photon is absorbed is $r_L$. To our knowledge, previous work that calculated evolution of $\Theta(x)$ and the step-size distribution ($P(r)$, see Eq. \ref{P_r}) considered an infinite sample for which $1-T=1$.\\

\begin{figure}
    \centering
    \includegraphics{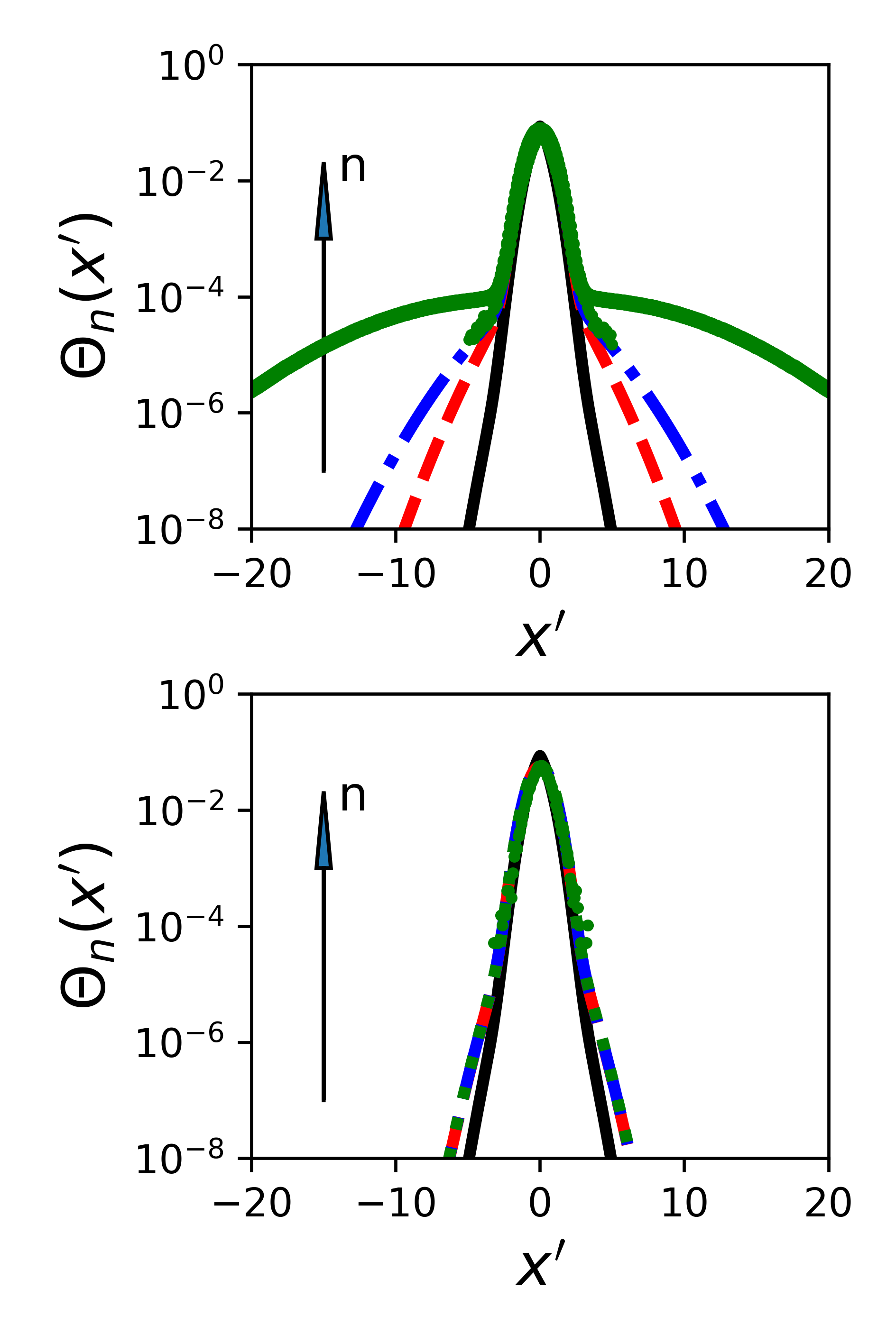}
    \caption{Calculated emission spectra for number of scattering events $n=1,5,10,50$ for the case R$_{II}$. (a) For infinite vapor; (b) for finite vapor with $r_L=100$.}
    \label{fig:Two_level_R2}
\end{figure}

\begin{figure}
    \centering
    \includegraphics{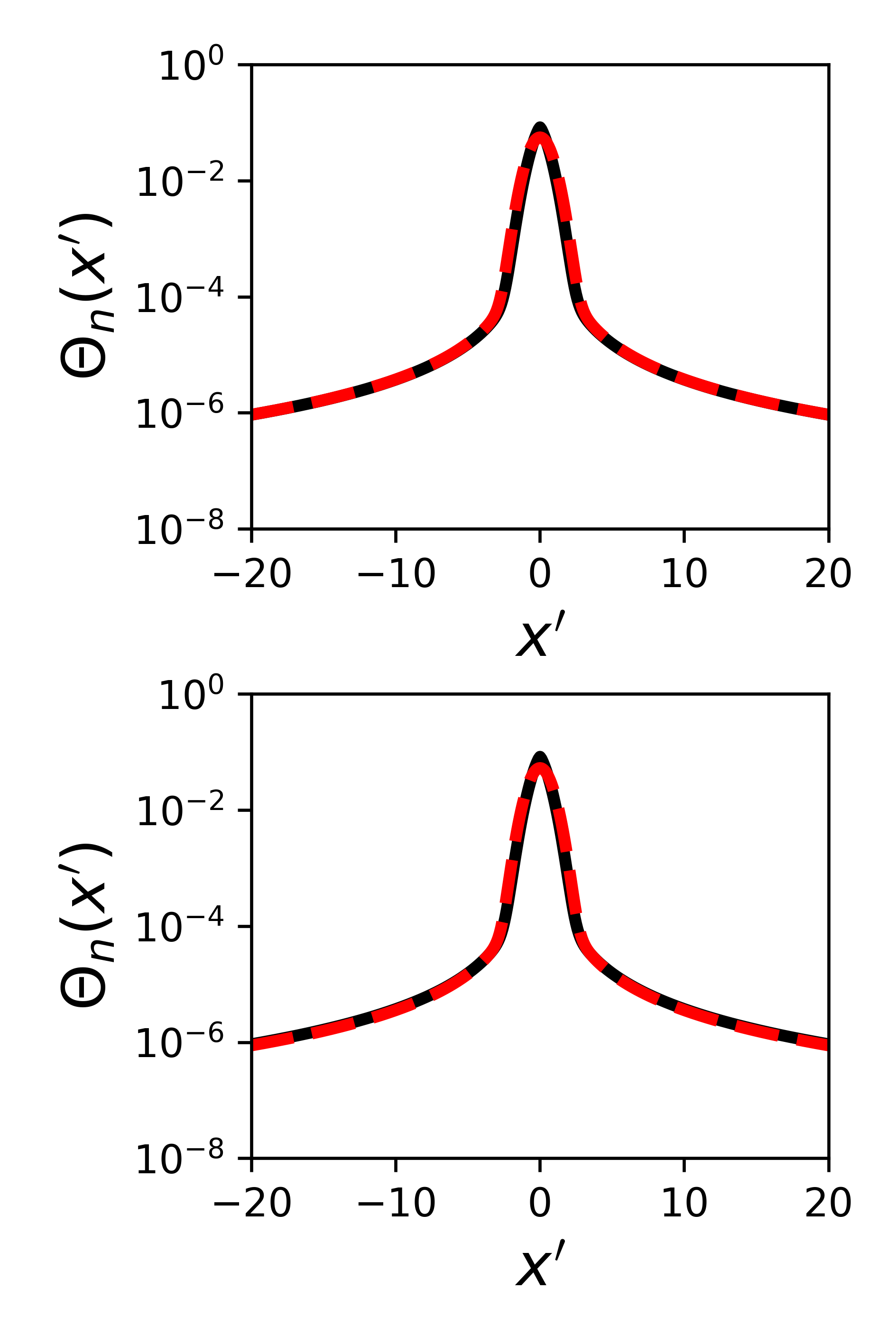}
    \caption{Calculated emission spectra for number of scattering events $n=1$ (black-solid line) and $10$ (red-dashed line) for the case R$_{III}$. (a) For infinite vapor; (b) for finite vapor with $r_L=100$.}
    \label{fig:Two_level_R3}
\end{figure}

In Figures \ref{fig:Two_level_R2} and \ref{fig:Two_level_R3} we plot the evolution of scattered spectra $\Theta_n$ with number of scattering events $n$, for the cases R$_{II}$ and R$_{III}$, respectively. In both figures the incident spectra are delta functions centered at the line-center $\Theta_0=\delta(0)$. In the R$_{II}$ case ( Fig. \ref{fig:Two_level_R2}(a)), for infinite vapor, we clearly see the formation of Doppler core and a truncation of emission in the wings \cite{Pereira2007,Lopez2023}. The wings are formed by off-resonance scattered photons in the atomic rest frame due to coherence of the scattering, with $x'\approx x$ within a Doppler width in the laboratory rest frame. This means that the frequency increment at successive scattering events is of a maximum of $\Gamma_D$. In contrast, for R$_{III}$ case, Fig. \ref{fig:Two_level_R3}(a), we observe formation of Lorentz wings since the first scattering event, as a consequence of redistribution of frequency in atomic frame (see Eq. \ref{lorentzian}), with CFR being reached after a few events in the laboratory frame. \\

For finite vapor, for R$_{II}$ case (see Fig. \ref{fig:Two_level_R2}(b)), the factor $\left[1-T\right]$ introduces a stronger cutoff in the emitted frequencies, as photons emitted far in the wings in step $n-1$ will not be scattered and will not participate in the line formation of step $n$. On the other hand, the factor $\left[1-T\right]$ does not impact the emission in R$_{III}$ case (Fig. \ref{fig:Two_level_R3}(b)), since emission in the atomic rest frame does not depend on incident frequency. In both Figs. \ref{fig:Two_level_R2}(a,b), we also plot scattered spectra obtained from Monte-Carlo simulations for infinite and finite vapor, respectively. We simulate the multiple scattering of photons from a two-level vapor and collect the frequency scattered and the step-size after each scattering event with the methods described in \cite{Carvalho2015}. In short, for each scattering event we draw the atomic velocity component parallel to the incoming photon \cite{Carvalho2015,Anderson1995}, and independently the velocity component perpendicular to the incoming photon. The incoming photon frequency is Doppler shifted to obtain the absorbed frequency in atomic rest frame. Then, we draw the direction of emitted frequency isotropically. As we simulate the R$_{II}$ case, the scattered frequency in the atomic rest frame is equal to the absorbed one. The scattered frequency is also Doppler shifted to obtain the frequency in the laboratory frame. The step-size is drawn from an exponential decay distribution (see Eq. \ref{P_Beer_Lambert}) and the process restarts for another photon. For Fig. \ref{fig:Two_level_R2}(a) no boundary was imposed for the vapor, while for results of Fig. \ref{fig:Two_level_R2}(b), at each step it was verified if the photon has escaped a cylindrical cell of radius $r_L$ and thickness $r_L$, with the outgoing photon being eliminated and a new photon being considered for the next scattering event. We see that the Monte Carlo simulations reproduce well the calculated spectra with Eq.~\ref{Eq:recurrence}, so this indicates that introducing the factor $[1-T(x)]$ in this equation is correct.


\begin{figure}
    \centering
    \includegraphics{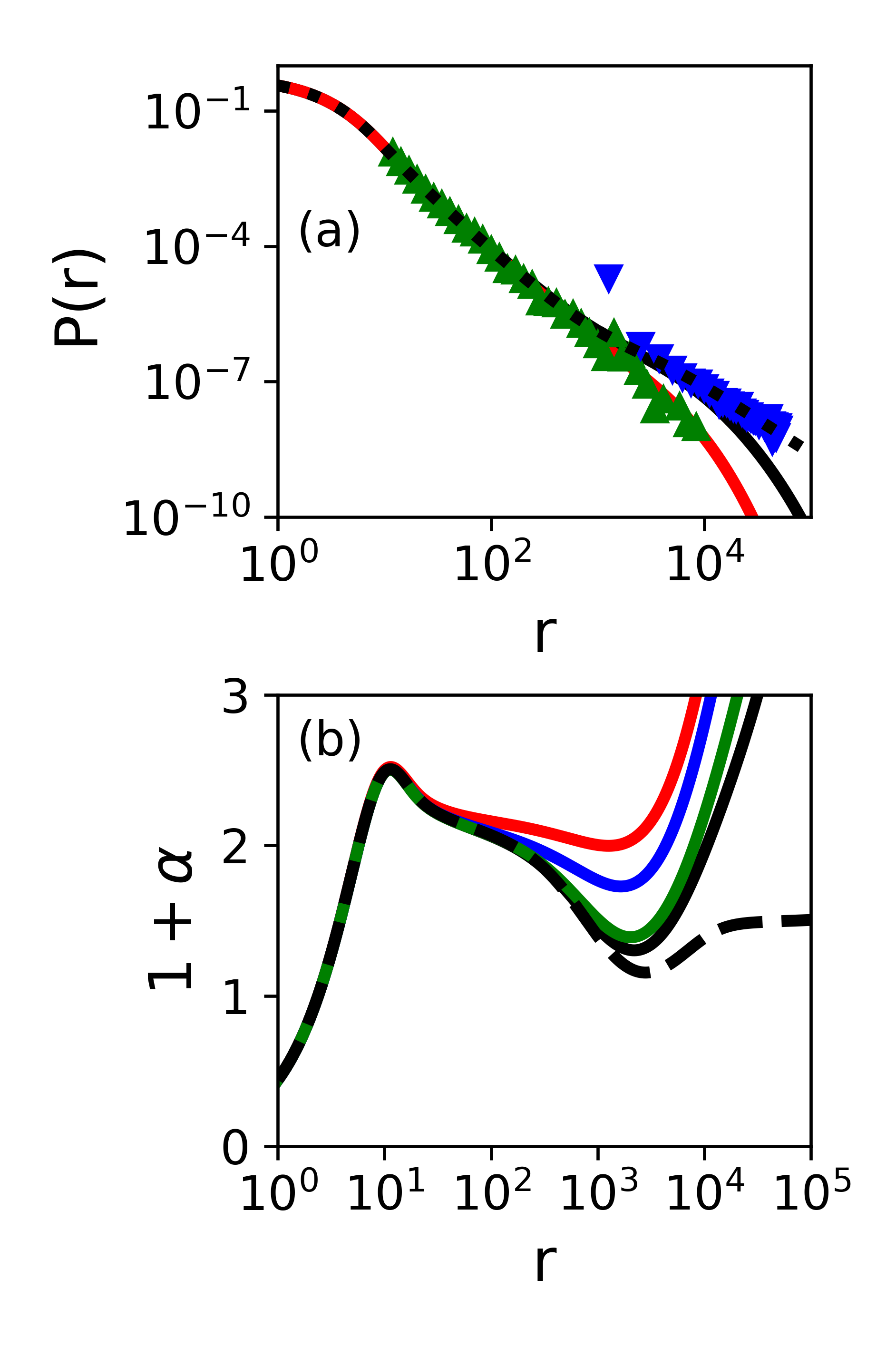}
    \caption{(a) Calculated step-size distribution $P(r)$ and (b) Lévy parameter obtained from $P(r)$; Black-solid: $R_{II}$ for infinite vapor; Black-dashed R$_{III}$ for infinite vapor; red-solid R$_{II}$ with $r_L=100$, blue-solid R$_{II}$ with $r_L=1000$, green-solid R$_{II}$ with $r_L=10^4$.}
    \label{fig:two_level_Passo}
\end{figure}

We plot in Fig. \ref{fig:two_level_Passo}(a) step-size distributions calculated with Eq. \ref{P_r} using the emission spectra for $n=10$. In Fig. \ref{fig:two_level_Passo}(b) we plot the step-size-dependent $\alpha$ value obtained as $1+\alpha=-\frac{d log P(r)}{dlog (r)}$. The black-solid lines are for R$_{II}$ case without the factor $\left[1-T\right]$ (infinite vapor). For moderate values of step-size ($r\sim 30 - 200$) the value of $\alpha$ converges towards $\alpha=1$ typical for CFR with Doppler profile \cite{Pereira2004,Pereira2007,Chevrollier2012}. Indeed, those step-size range are done typically by photons emitted in the Doppler core of $R_{II}$ emission spectra \cite{Macedo2021}. After a certain step-size ($r\sim 10^4$), we observe a cutoff in the $P(r)$, due to cutoff of the emission frequency, that leads to an increase of the $\alpha$ value, losing the Lévy flight character. For the R$_{III}$ case (dashed-black lines), the step-size distribution is similar to the R$_{II}$ for short jumps as scaterred profile in $R_{III}$ is a Voigt profile with a Doppler core. For long jumps $\alpha\rightarrow 0.5$ as a result of contribution of the Lorentz wings \cite{Pereira2004,Chevrollier2012,Macedo2021}. For finite vapor the earlier cutoff in frequency for the R$_{II}$ case (Fig. \ref{fig:Two_level_R2}(b)) shifts the cutoff in the step-size to lower values, resulting in larger $\alpha$ values relative to infinite vapor. For instance, we observe that for infinite vapor, the $\alpha$ value for R$_{II}$ attains values as low as $\alpha=0.3$ (see black-solid line in Fig. \ref{fig:two_level_Passo}), close to R$_{III}$ case (black-dashed line), whereas, for vapor of small size ($r_l\sim 100$, red-solid line) those low values are not attained. Note also that for higher $r_L$ the $1+\alpha$ values approaches the infinite vapor dependence with cutoff occurring for higher step-size values. The change of the scattered spectra, of the step-size distribution and of the $\alpha(r)$ value as a function of the system size was not, to our knowledge, discussed before and is one of the main results of this work. \\

\begin{figure}
    \centering
    \includegraphics{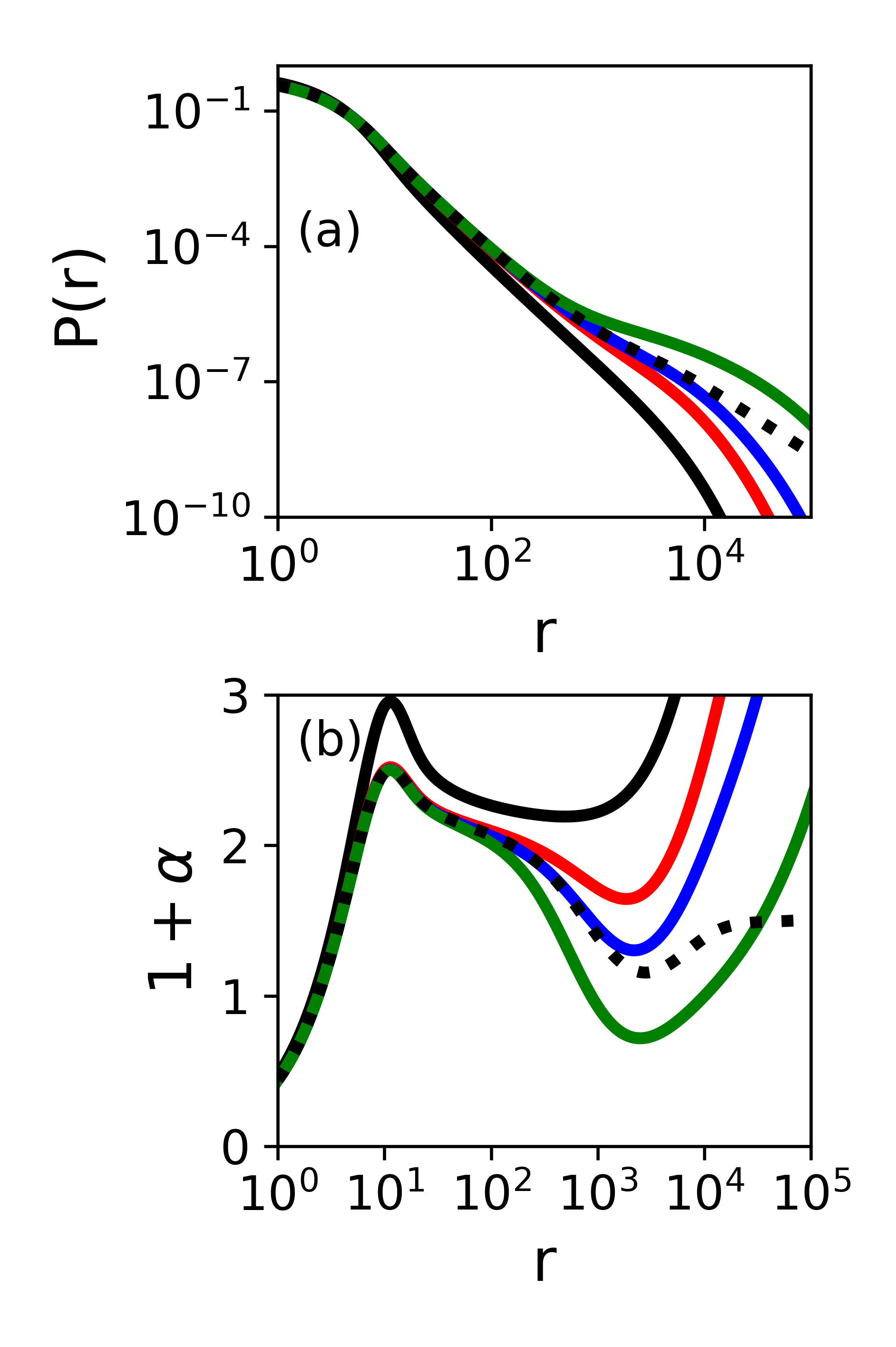}
    \caption{(a) Calculated step-size distribution $P(r)$ and (b) Lévy parameter obtained from $P(r)$ for infinite vapor and diferent number of scaterring events; Black-dashed for R$_{III}$ case and $n=10$; black-solid, red-solid, blue-solid and green-solid lines for R$_{II}$ case with $n=1,5,10,50$, respectively.}
    \label{fig:two_level_Passo_R2}
\end{figure}

\section{Step-length distribution for alkali vapor}
Experimental realization of Lévy flight of photon were performed with alkali vapors \cite{Mercadier2009,Mercadier2013,Baudouin2014,Araujo2021,Macedo2021,Lopez2023}, Rb and Cs. Rb and Cs vapors are not two-level systems, instead, they exhibit excited hyperfine levels that in general are not resolved at room temperature due to Doppler broadening (for D$_2$ lines of Rb and Cs). Moreover, optical pumping to initially not-coupled hyperfine levels may occur, changing the effective system size. In this section we calculate the step-size distribution for Cs D$_2$ and D$_1$ lines, as well as, for Rb D$_2$ line and connect the obtained $\alpha(r)$ values to previous experimental results.\\

To take in account the excited and ground hyperfine manyfolds the redistribution function $R(x,x')$ is summed over all possible transitions. The photon may be absorbed in the transition $F_1\rightarrow F'$ and then scattered in transition $F'\rightarrow F_2$, with $F_1$ and $F_2$ denoting, respectively, initial and final hyperfine ground levels and $F'$ denoting excited hyperfine level. The redistribution function is then calculated as \cite{Mercadier2013}
\begin{equation}
\begin{split}
    R(x,x')=\sum^{F_M}_{F_1=F_m}\sum^{F_M}_{F_2=F_m}\sum^{F'_M}_{F'=F'_m}\frac{2F_1+1}{\sum^{F_M}_{F=F_m}\left(2F+1\right)}\frac{S_{F_1F'}S_{F_2F'}}{\sum^{F_M}_{F=F_m}S_{FF'}}\\ \times R\left(x-x_{F_1F'},x'-x_{F_2F'}\right), \label{EqRedistribution}
\end{split}
\end{equation}
with $F_m$ ($F'_m$) and $F_M$ ($F'_M$) denoting the minimum and maximum value of ground (excited) $F$ quantum number, respectively. $S_{FF'}$ are factors to take in account relative strengh of the transitions (calculated from the Clebsch-Gordan and the Wigner 3-j coefficients) \cite{Steck2003} and $x_{FF'}$ are the normalized frequency of the transition $F\rightarrow F'$.  The normalized absorption profile is generalized to:
\begin{equation}
    \phi_T(x)=\sum^{F_M}_{F=F_m}\sum^{F'_M}_{F'=F'_m}\frac{2F+1}{\sum^{F_M}_{F=F_m}\left(2F+1\right)}S_{FF'}\phi(x-x_{FF'}).
\end{equation}

\subsection{Cesium D$_2$ line}
\begin{figure}
    \centering
    \includegraphics{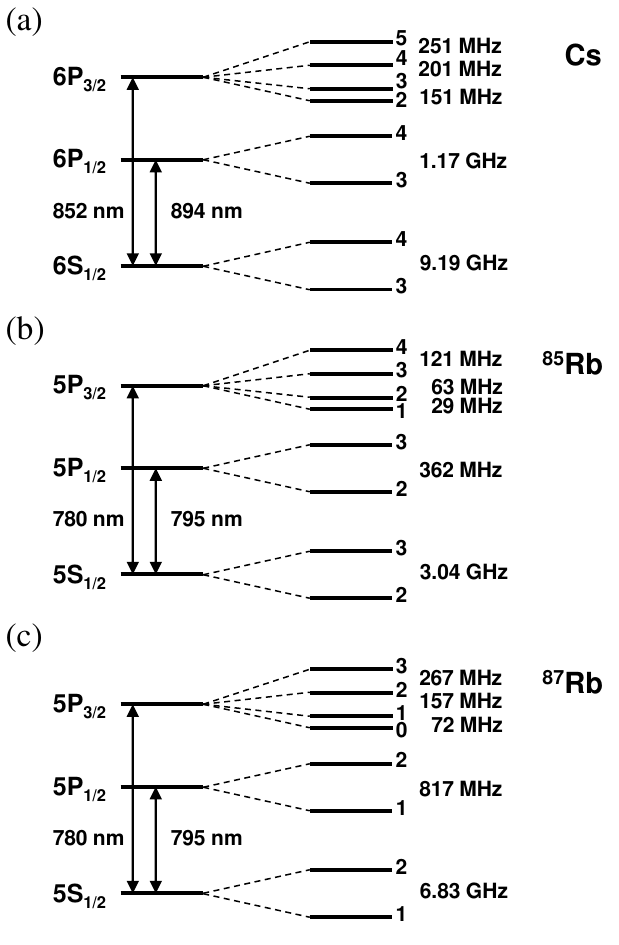}
    \caption{ Schematics of the D1 and D2 lines of (a) Cs, (b) $^{85}$Rb and (c) $^{87}$Rb (out of scale).}
    \label{fig:Cs-Rb}
\end{figure}

The Cesium D$_2$ transition corresponds to transition from $6S_{1/2}$ ground level to $6P_{3/2}$ excited level (see Fig.\ref{fig:Cs-Rb}), with a wavelength of $\lambda_{D2}=852$ nm. Ground manyfold has two hyperfine levels $F=3$ and $F=4$ separated by atomic clock frequency of $x_{Clock}=9.193$ GHz/$\Gamma_D$. Excited manifold has hyperfine levels $F'=2-5$, with separations of $151$ MHz, $201$ MHz,  and $251$ MHz from the lower to the upper level. Note that separation between excited level is of the same order of the Doppler-half-width at 1/e level $\Gamma_D=u/\lambda=254$ MHz at $100^\circ$C used for the calculations, with $u$ the most probable atomic speed.  \\

\begin{figure}
    \centering
    \includegraphics{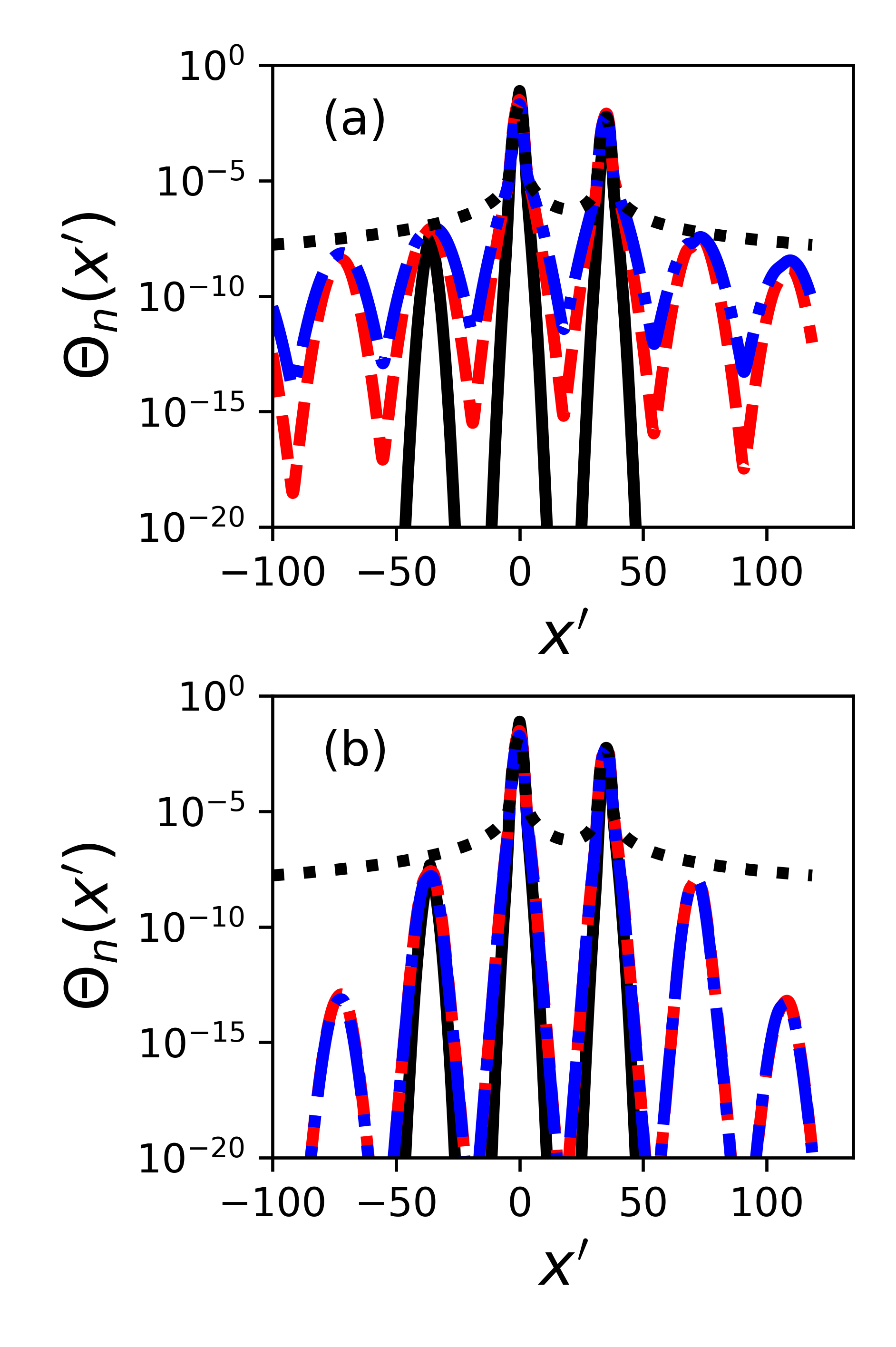}
    \caption{Frequency redistribution for Cs D$_2$ line for different numbers of scattering events: (a) for infinite vapor; (b) for finite vapor with $r_L=100$. Black-solid , red-dashed and blue-dotted-dashed lines are for R$_{II}$ case after scattering event of number n = 1,5 and 10, respectively. Black-dotted line is for R$_{III}$ case after scattering event number n=10.}
    \label{fig:Freq_D2}
\end{figure}

\begin{figure}
    \centering
    \includegraphics{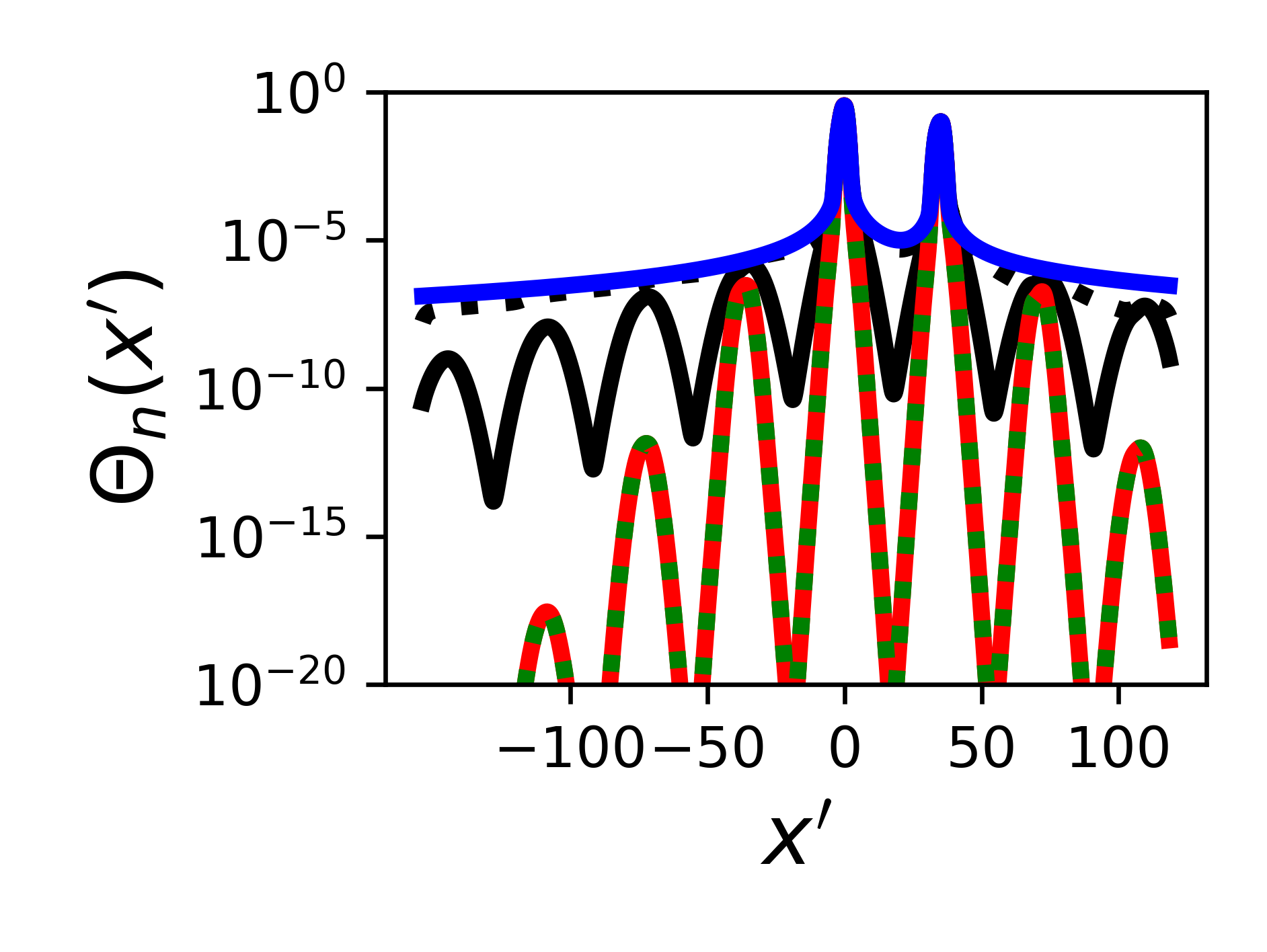}
    \caption{Comparison of frequency redistribution for Cs D$_2$ line for Lorentzian and Dirac delta incident spectra for the R$_{II}$ case with n=10. Black-solid and red-dotted lines are for incident $\Theta_0=\delta(x)$ for infinite and finite vapor ($r_L=100$), respectively. Black-dotted and green-dotted are for incident Lorentzian spectrum for infinite and finite vapor ($r_L=100$), respectively. The blue-solid line is for the R$_{III}$ case.}
    \label{fig:Freq_D2_2}
\end{figure}

\begin{figure}
    \centering
    \includegraphics{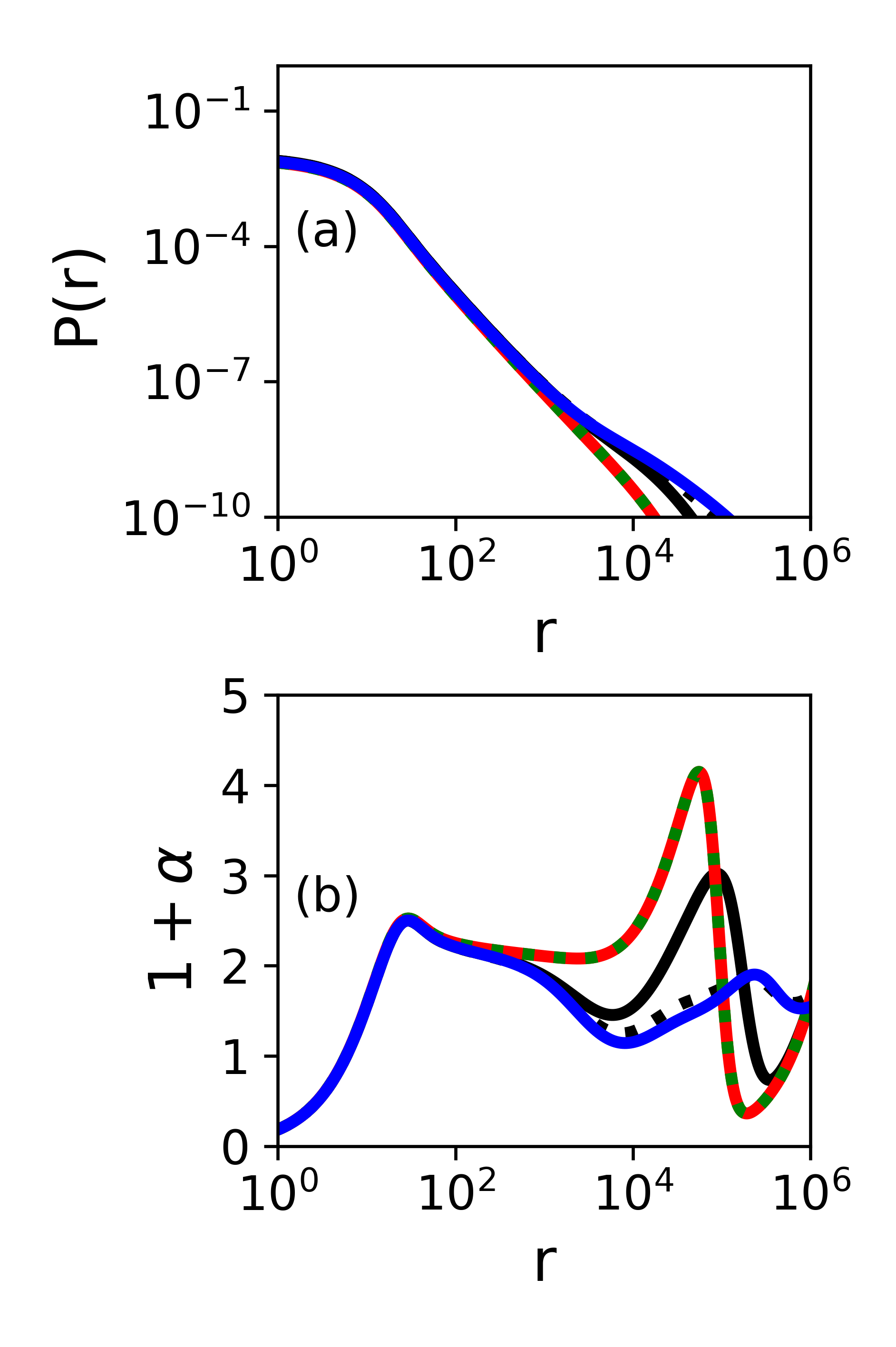}
    \caption{(a) Step-size distribution for Cs D$_2$ line. (b) Exponent $1+\alpha$ calculated from the step-size distribution, for the R$_{II}$ case with $n=10$. Black-solid and red-dotted lines are for incident spectrum $\Theta_0=\delta(x)$ for infinite and finite vapor ($r_L=100$), respectively. Black-dotted and green-dotted lines are for incident Lorentzian spectrum for infinite and finite vapor ($r_L=100$), respectively. The blue-solid line is for the R$_{III}$ case.}
    \label{fig:Passo_D2}
\end{figure}

In Fig. \ref{fig:Freq_D2} we show the scattered spectra $\Theta_n$ in the R$_{II}$ case for incident photon at the cycling transition $F=4\rightarrow F'=5$ (taken as $x=0$). For both infinite (Fig. \ref{fig:Freq_D2}(a)) and finite vapor (Fig. \ref{fig:Freq_D2}(b)) we observe that, after the first scattering event three peaks appears: (i) a main peak around incident detuning $x=0$: absorption occurs in $F=4\rightarrow F'=3,4,5$ and is scattered for the original ground state. The peak around $x=0$ is broadened due to Doppler broadening; (ii) a peak around ($x_{F=3,F'}=x_{Clock}\sim 36.2$): the photons that are absorbed in $F=4\rightarrow F'=3,4$ transition might decay to $F=3$ (optical pumping) being responsible for the peak to the right around $x_{Clock}$; (iii) incident photon may also be absorbed in the $F=3\rightarrow F'=2,3,4$ transitions, although with a much lower probability. Those photons are absorbed detuned by $\sim x_{Clock}$ in the atomic rest frame and are scattered also detuned by $x_{Clock}$ (coherence in the atomic rest frame). If they are scattered towards ground level $F=3$ they will sum up to the peak at $x=0$. However, if the photons are scattered toward ground level $F=4$ they result in the peak at the left around $-x_{Clock}$ from $F=4$.\\

For subsequent scattering events ($n>1$) others peaks form spaced by $x_{Clock}$ following a similar mechanism. Two new peaks, one to the left and one to the right, spaced by $x_{Clock}$ from previous one, form at each new scattering event. For an infinite vapor (Fig. \ref{fig:Freq_D2}(a)), the amplitude of the peaks follows the absorption profile, with a Lorentzian decay at the wings. For finite vapor (see Fig. \ref{fig:Freq_D2}(b)) the probability that a photon scattered at the wings to be scattered again by the vapor is low and the amplitude of the peaks decay stronger than the Lorentzian wing. For the R$_{III}$ case, similar results are obtained for infinite and finite vapor (see Figs. \ref{fig:Freq_D2}(a,b)) and the lateral peaks does not appear since they are caused by coherent scattering in the atomic rest frame that are absent in R$_{III}$ case. 

We have also calculated scattered spectra considering an incident spectra $\Theta_0$ with Lorentzian profile of width 10 MHz, typical for experiments with diode lasers \cite{Mercadier2013}. We plot calculated spectra for $n=10$ in dotted-lines in Fig. \ref{fig:Freq_D2_2} and compare them with previously calculated for $\Theta_0=\delta(x)$ (solid lines).  For infinite vapor and R$_{II}$ case (black-dashed lines in Fig. \ref{fig:Freq_D2}(a)), far-from-resonance photons will be scattered also far from resonance, making the Lorentzian wings of the incident spectra to survive to subsequent scattering events.  For finite vapor in the R$_{II}$ case with incident Lorentzian profile (green-dotted line) no changes are observed with respect to incident delta profile (red-solid line) since photons scattered at the wings are eliminated from subsequent scattering due to the $[1-T]$ factor. In Fig. \ref{fig:Freq_D2_2} is also shown the $R_{III}$ case (blue-solid line) for comparison. The scattered spectra for R$_{III}$ case does not depend on incident spectra since there is no coherence in atomic rest frame.

In Fig. \ref{fig:Passo_D2}(a) we plot the obtained step-size distribution for $n=10$ for R$_{II}$ case in both infinite and finite vapor and for the R$_{III}$ case (for the $R_{III}$ case, the same results are obtained for finite and infinite vapor). The corresponding $1+\alpha$ values are plotted in Fig. \ref{fig:Passo_D2}(b). Values of $1+\alpha$ are similar to those obtained for two-level model for $r<10^4$ \footnote{Note that transition from $\alpha\sim 1$, contribution of Doppler core, to $\alpha\sim 0.5$, contribution of Lorentzian wings, occurs for higher $r$ values in D$_2$ compared to two-level system. This is because for the same optical density $N\sigma(0)L$, $r$ for D$_2$ line is almost the double of that for two-level system since there is a ratio of $\sim 2$ for $\phi(0)$ between the two cases.}. For $r\sim 10^4$ an oscillation of $1+\alpha$ values appears due to the presence of the second ground state peak. The appearance of the lateral peaks in R$_{II}$ case contributes to an increasing of the amplitude of the wings of $\theta_n$ for increasing $n$, similar to the increasing of the level of the Lorentzian wings for two-level case. This increase of $\Theta_n$ in the wings makes step-size distribution and $1+\alpha$ values for R$_{II}$ with infinite vapor (black-solid line, with $\Theta_0=\delta(x)$) approach the R$_{III}$ (blue-solid line). For incident Lorentzian profile (black dotted-line) $1+\alpha$ curve is very close to that of the R$_{III}$ case since scattered spectra preserve the heavy-tailed wings (see Fig. \ref{fig:Freq_D2_2}). $1+\alpha$ value for finite vapor finite vapor does not depend on the incident spectra (solid-red line for $\Theta_0=\delta(x)$ and green-dotted line for Lorentzian profile) as they have the same scattered spectra (see Fig. \ref{fig:Freq_D2_2} and discussion above).

Measurement of $\alpha$ value for Cs D$_2$ line was reported in \cite{Macedo2021,Lopez2023} with values ranging from $\alpha\sim 0.8$ for $N\sigma(0)L\sim 10^2$ and thus $r\sim 400$ to $\alpha\sim 0.6$ for $N\sigma(0)L\sim 10^4$ and thus $r\sim 4\cdot 10^4$ \cite{Lopez2023}. Although a direct comparison with the present calculations is not straightforward since parameters such as $\Gamma$ and $\Gamma_D$ in \cite{Lopez2023} (which changes for each value of $r$) are different than those used here, these measurements seem to be consistent with our calculations. 


\begin{figure}
    \centering
    \includegraphics{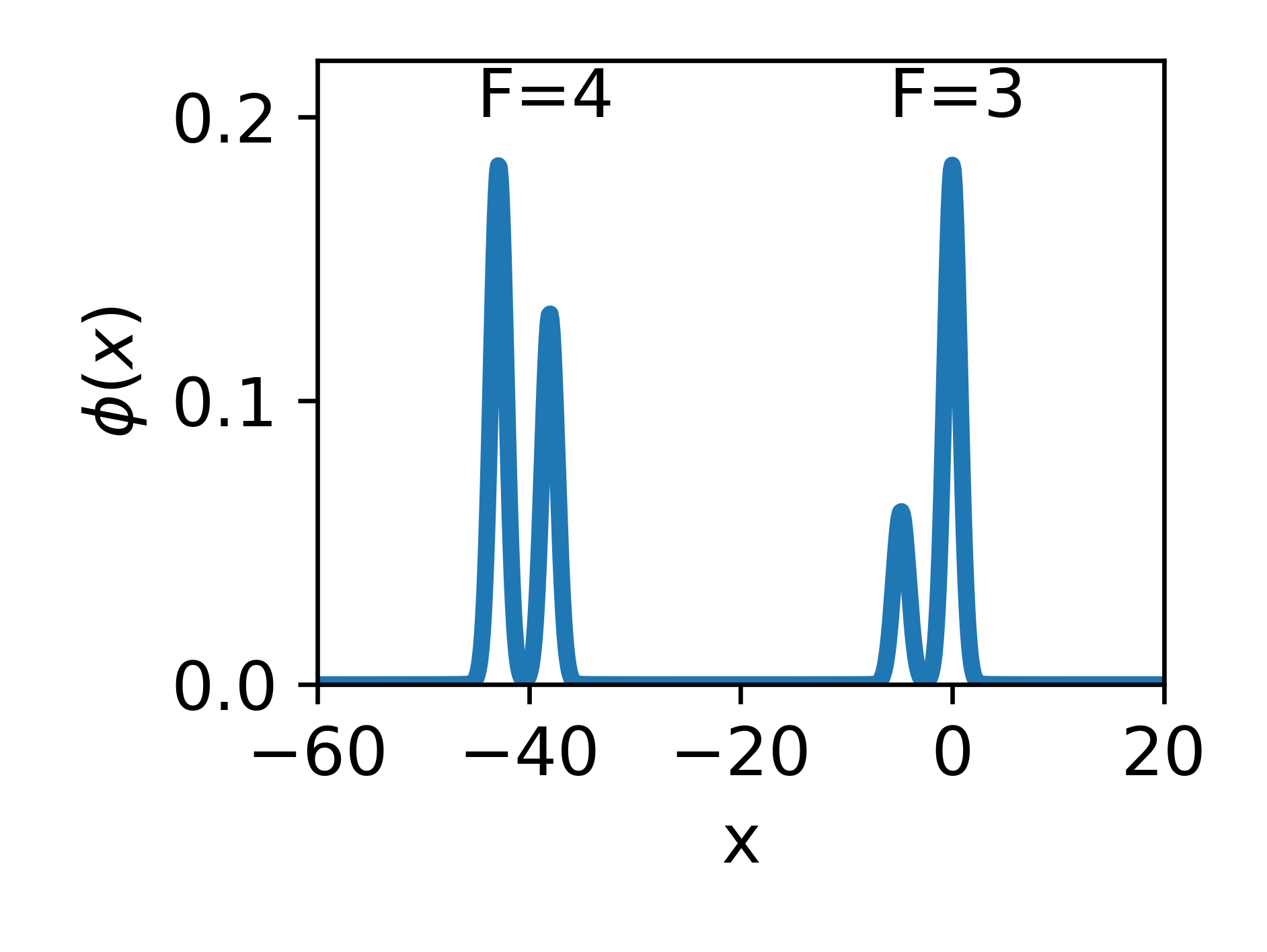}
    \caption{Normalized absorption coefficient $\phi(x)$ for the cs D$_1$ line.}
    \label{fig:Coef_D1}
\end{figure}
\subsection{Cesium D$_1$ line}
\begin{figure}
    \centering
    \includegraphics{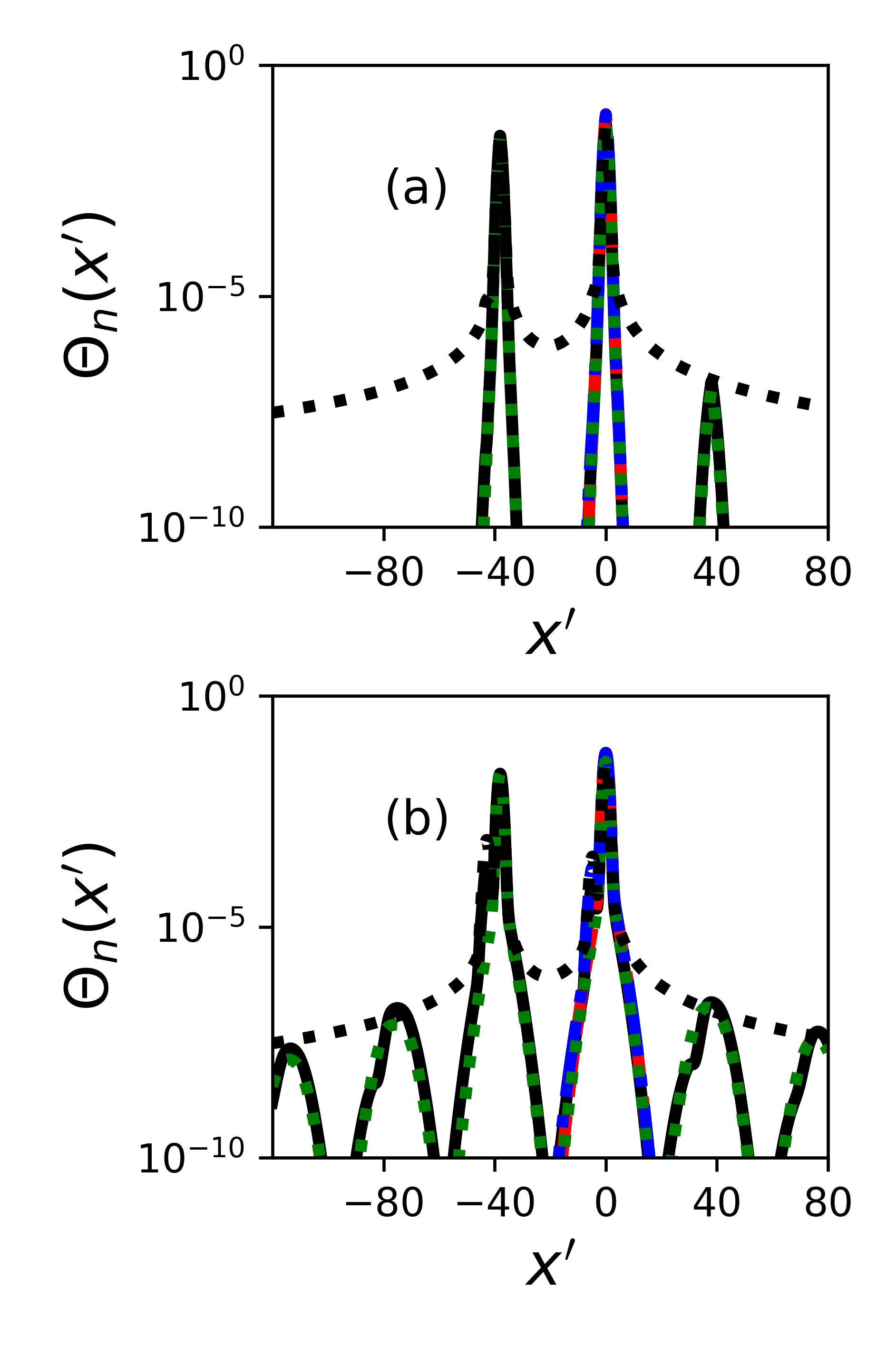}
    \caption{Scattered spectrum after the first scattering event for Cs D$_1$ line for the R$_{II}$ case (black-solid line). Red-dashed lines: calculations considering only the $F=3\rightarrow F'=4$ transition; Blue-dashed-dotted line calculations considering only $F=3\rightarrow F'=3,4$ transitions; Green-dotted calculations considering only $F=3,4\rightarrow F'=4$ transitions. Black-dotted line: calculations for the R$_{III}$ case. (a) For first scattering event $n=1$; (b) For $n=10$ scattering event.}
    \label{fig:Theta1_D1}
\end{figure}

\begin{figure}
    \centering
    \includegraphics{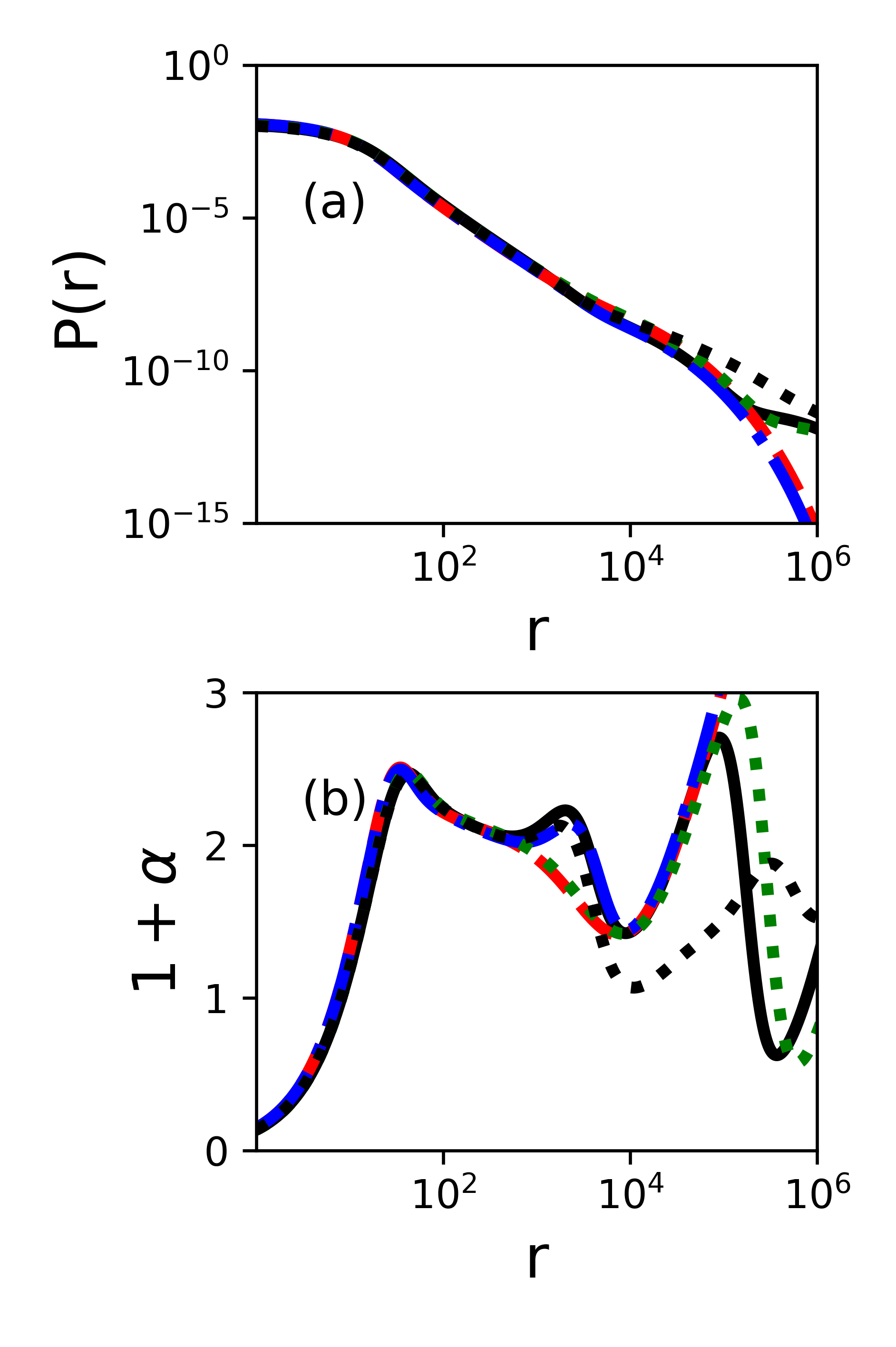}
    \caption{Step-size distribution after the scattering event $n=10$ for Cs D$_1$ line (black-solid line). Red-solid lines: calculations considering only the $F=3\rightarrow F'=4$ transition; Blue-solid calculations considering only $F=3\rightarrow F'=3,4$ transitions; Green-solid calculations considering only $F=3,4\rightarrow F'=4$ transitions.}
    \label{fig:Step_D1}
\end{figure}

The Cesium D$_1$ transition corresponds to transition from $6S_{1/2}$ ground level to $6P_{1/2}$ excited level (see Fig. \ref{fig:Cs-Rb}, with a wavelentgh of $\lambda_{D1}=894$ nm. Ground manifold has two hyperfine levels $F=3$ and $F=4$ separated by atomic clock frequency $x_{Clock}=38$ \footnote{The value of $x_{Clock}$ is a little higher here relative to D$_2$ line as $\lambda_{D1}>\lambda_{D2}$. We have kept in the calculations same vapor temperature of $T=100^\circ$C, resulting in $\Gamma_D=242$ MHz and almost unchanged Voigt parameter $a=10^{-2}$. }. Excited manyfold has hyperfine levels $F'=3,4$, with separations of $x_{F'}=(1168\,\mathrm{MHz})/\Gamma_D=4.8$, larger than Doppler width. The excited hyperfine manyfold is thus well resolved.\

The normalized absorption profile is shown in Fig. \ref{fig:Coef_D1}. One can see two pairs of peaks with each pair corresponding to a given ground hyperfine level. As stated above the excited levels are well resolved since separation is of the order of $x_{F'}$.

In this section we will limit our discussion for the case of infinite vapor, as the influence of the finite size of the vapor gives qualitatively similar results to those discussed for the D$_2$ line and the two-level system. Three pairs of peaks appear for the scattered spectrum after first scattering event (see Fig. \ref{fig:Theta1_D1}(a), black-solid line), with the same origin of the discussed peaks for D$_2$ line. In Fig. \ref{fig:Theta1_D1}(b)  scattered spectrum after $n=10$ is shown exhibiting lateral peaks similar to those obtained in D$_2$ line. In Fig. \ref{fig:Step_D1}  we show step-size distribution and expected $1+\alpha$ parameter obtained for the D$_1$ line. Oscillations are observed in the dependence of $1+\alpha$ value with step-size. To understand the origin of those oscillations we have calculated again the redistribution function keeping only selected terms in $F_1$ , $F_2$ and $F'$ in Eq. \ref{EqRedistribution}. First, we keep only the $F=3\rightarrow F'=4
$ transition and retrieve results for a two-level model (see red-solid lines in Fig. \ref{fig:Theta1_D1}). 
In blue, we show calculations considering only one ground state and both excited levels (see Fig. \ref{fig:Theta1_D1}). Oscillations in emission profile does not appear. We see that the existence of the resolved second excited level is responsible for oscillations of $1+\alpha$ value in the range $r\sim 10^3-10^4$. Photons scattered at the transition between Doppler core and Lorentz wings are responsible for steps in the range $r\sim 10^3-10^4$ for two-level model. The presence of the second excited level makes steps emitted in this frequency range smaller than for two-levels model, increasing the decay of $P(r)$ in this range. In green, we consider only one excited level and both ground levels. Emission spectrum and $1+\alpha$ dependence on step-size resembles to the results of the $D_2$ line, with oscillation appearing in $1+\alpha$ for larger steps $r\sim10^5-10^6$ due to the presence of the second ground state. Although the Cs D$_1$ line seems to be a interesting system to explore the dependence of the $\alpha$ value with step-size, to our knowledge no measurement related with Lévy flight were done in this transition.

\subsection{Rb D$_2$ line}
The rubidium D$_2$ corresponds to transition from ground $5S_{1/2}$ to excited $5P_{3/2}$ levels. Glass cells used in experiments usually contains two Rb isotopes with natural abundance of $72.17\%$ of $^{85}$Rb and $27.83\%$ of $^{87}$Rb \cite{Steck2003}. $^{85}$Rb has two ground states $F=2, 3$ separated by $3036$ MHz and four excited hyperfine levels $F'=1-4$ separated by $29.4$ MHz, $63.4$ MHz and $120.6$ MHz. $^{87}$Rb has two ground states $F=1,2$ separated by $6835$ MHz and four excited hyperfine levels $F'=0-3$ separated by $72.2$ MHz, $156.9$ MHz and $266.6$ MHz. \\

Measurements of Lévy $\alpha$ parameter in Rb vapor were performed in \cite{Mercadier2009,Mercadier2013} by observing directly the step-size distribution, i.e., the positions of the photon first scattering event were recorded with a CCD camera. In order to avoid multiple scattering, the $P(r)$ was measured for small system size, at maximum $r_L\sim 20$, and the Lévy parameter was measured to be $\alpha=1.09\pm 0.15$ \cite{Mercadier2009,Mercadier2013}. The authors compared measured value with different level-scheme models including the whole hyperfine manyfold. Measurement of Lévy parameter for Rb vapor were also done in the multiscattering regime \cite{Baudouin2014} obtaining an $\alpha=1.03\pm0.15$ for system size up to $r_L\sim 10^5$ \footnote{The authors have used different methods to measure $\alpha$ in \cite{Baudouin2014}. The system size of $r_L\sim 10^5$ reported here is taking from measurements using radial profile of scattered light. In those measurements step-sizes up to 3 cm were observed. Note that the authors in \cite{Baudouin2014} reports results as a function of (longitudinal) opacities corresponding to a longitudinal thickness of the sample of 5 mm. The transversal opacity (corresponding to step-size of 3 cm) is 6 times larger than the reported longitudinal opacity. The relation between dimensionless system size and opacity used in \cite{Baudouin2014} is $r_L=O\pi a$, with $O$ the opacity.}. \\

For calculating the scattered spectrum and the step-size distribution we take in account both isotopes by $R(x,x')=N^{(85)}R^{(85)}(x,x')+N^{(87)}R^{(87)}(x,x')$, with $N^{(i)}$ natural abundance of isotope $^{(i)}Rb$ and $R^{(i)}$ the redistribution function by summing over hyperfine transitions for isotope $^{(i)}Rb$. Similarly, $\phi_T(x)=N^{(85)}\phi^{(85)}_T(x)+N^{(87)}\phi^{(87)}_T(x)$. To compare our calculations to the results of \cite{Baudouin2014}, we consider a laser with Lorentzian spectral distribution, width of $\gamma= 2$ MHz and at crossover resonance between $F=2\rightarrow F'=2,4$ of $^{85}Rb$. 

\begin{figure}
    \centering
    \includegraphics{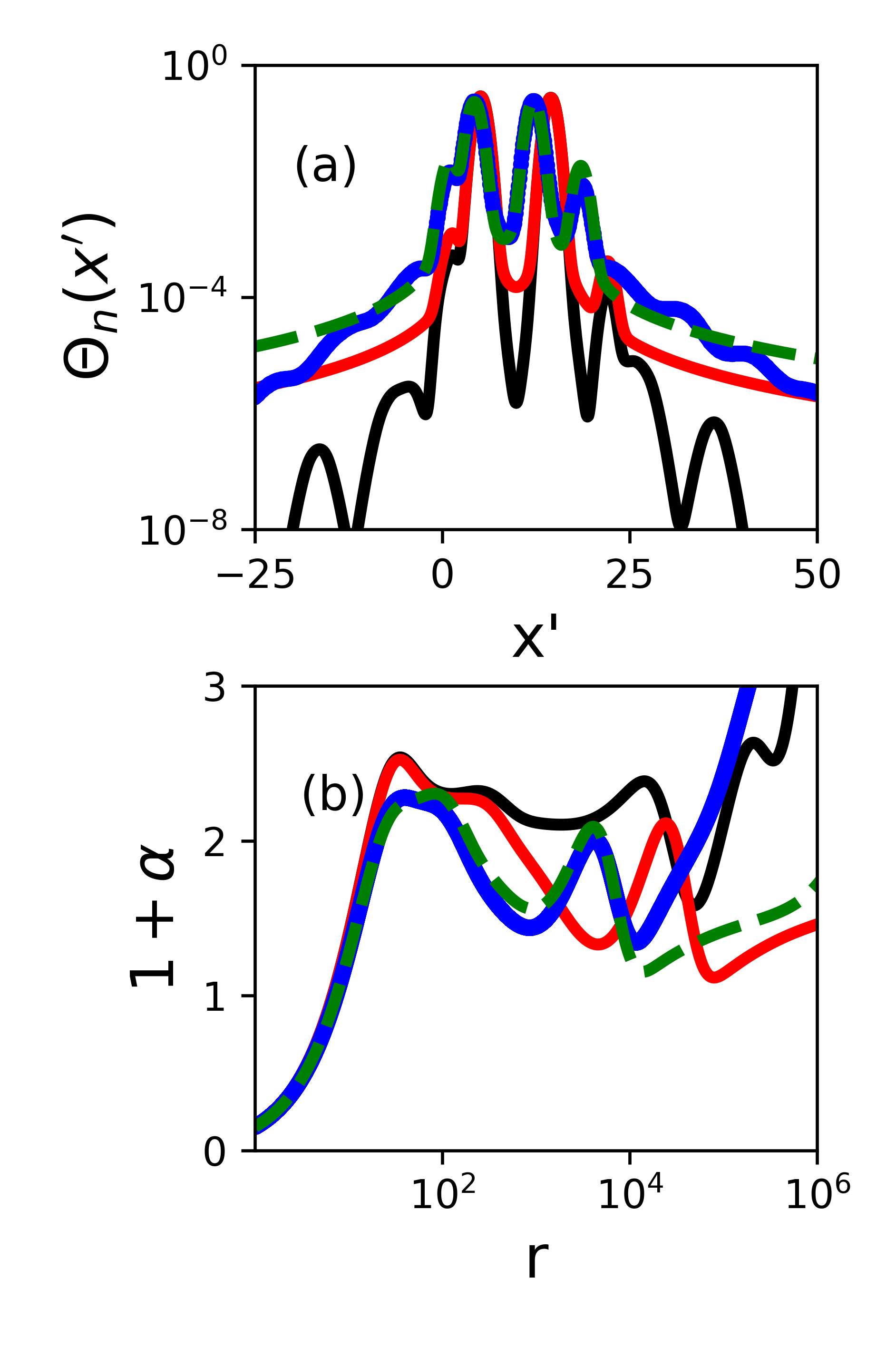}
    \caption{(a) Spectrum of scattered light after scattering event $n=10$ for Rb D$_2$ line. (b) $1+\alpha$ after scattering event $n=10$ for Rb D$_2$ line. Black: R$_{II}$ case and $N_{Low}$ and $r_L=100$; Black-dotted : R$_{II}$ case and $N_{High}$ and $r_L=10^4$; Red : R$_{III}$ case and $N_{Low}$ and $r_L=100$; Blue: : R$_{III}$ case and $N_{High}$ and $r_L=10^4$.}
    \label{fig:Rb_Redist}
\end{figure}

The scattered spectra after $n=10$ events are shown in Fig. \ref{fig:Rb_Redist}(a) exhibiting oscillations for the $R_{II}$ case and Lorentzian wings for $R_{III}$ case. As transitions from $^{87}Rb$ $F=2$ and from $^{85}Rb$ $F=3$ are close to each other the excitation can pass from one isotope to the other and the four peaks appears in the scattered spectrum. In this sense, scattered spectrum is similar to the one of the D$_1$ Cs line. In Fig. \ref{fig:Rb_Redist}(b) we show the calculated $1+\alpha$ value as a function of dimensionless distance $r$ for two different vapor temperatures. In the calculations we consider that the temperature of Rb liquid in equilibrium with the vapor is the same as the temperature of the vapor, so the temperature affect both vapor density and total homogeneous width (Doppler width is almost unaffected). The two densities are: i) $N_{Low}=5\times10^{16}$ atoms/m$^3$, with homogeneous width of $\Gamma^{(85)}=6.07$ MHz (no collisional broadening) and Voigt parameter $a=10^{-2}$. For the low density we use a system size of $r_L=20$ in the calculations. ii) $N_{High}=2.5\times10^{20}$ atoms/m$^3$, with homogeneous width of $\Gamma=4.5\Gamma_n$ and Voigt parameter $a=3.5\times 10^{-2}$. For the high density we use a system size of $r_L=10^5$ in the calculations. For low density, we obtain $\alpha$ slightly over 1 for a range of dimensionless parameter $r\sim 10-10^4$. This is consistent both with results of \cite{Mercadier2009,Mercadier2013}. An oscillation of $\alpha$ value in the range  $r\sim 10^4-10^5$ is observed due to the presence of the many peaks of scattered profile. The dependence of $1+\alpha$ with $r$ is similar for high density step-size distribution with structures shrunk to lower $r$ value, due to increase of Voigt parameter \cite{Chevrollier2012}. This is due for the relative probability of absorption of a photon scattered at the wings relative to one scattered at the core increases with density as Voigt parameter is higher. The high density used in calculations is close to the one used for higher opacity in \cite{Baudouin2014} ($r_L\sim 10^5$) for which $\alpha<1$ was measured. Indeed, our calculations are consistent with measurement of \cite{Baudouin2014} with a decrease of $\alpha$ around $r\sim 10^5$. 

\section{Conclusions}
We have calculated the frequency redistribution and the step-size distribution for light diffusion in resonant atomic vapor, where photon random walks can be described at Lévy flights \cite{Pereira2004}. Describing locally the step-size distribution as $P(r)\propto r^{-1-\alpha}$ we extract the Lévy parameter $\alpha$ as a function of $r$. Step-size distribution at step $n$ depends on the scattered spectrum at step $n$, that depends itself on the scattered spectrum at step $n-1$ for Partial Frequency Redistribution \cite{Pereira2004,Mercadier2013,Lopez2023}.\\

First, we have considered the case of a vapor of two-level atoms. We have introduced the finite size of the system in the calculations of the scattered spectrum at step $n$. For the R$_{II}$ case, the finite size of the system introduces a cutoff in scattered spectrum that is stronger than the usual cutoff considered for infinite vapor \cite{Pereira2007}. This stronger cutoff implies a cutoff in step-size changing the expected $\alpha$ values relative to infinite vapor. As expected, we have verified that no change is introduced in the R$_{III}$ case due to finite size of the system. Also, we have showed that, for infinite vapor, the wings of the R$_{II}$ case approaches progressively the Lorentzian wings of the R$_{III}$ case for increasing number of scattering events $n$. This fact may have an impact on very large systems such as astrophysical systems. Second, we have calculated the scattered spectrum and step-size distribution for alkali vapors, specifically Cs D$_1$ and D$_2$ lines and Rb D$_2$ line. We showed that the typical multilevel structure of alkali atoms introduces oscillations in the scattered spectra for the R$_{II}$ case. Also, oscillations on the L\'evy exponent $\alpha$ value are expected from step-size distributions due the existence of several peaks in the scattered spectrum. These peaks are related to the existence of two hyperfine ground states of alkali atoms, to the resolved hyperfine excited levels of the Cs D$_1$ line and to the two isotopes of Rb D$_2$ line. We have found that the calculated $\alpha$ values using the model described in this work are consistent with previously measured $\alpha$ parameter for Cs and Rb D$_2$ lines \cite{Mercadier2009,Mercadier2013,Macedo2021,Lopez2023}.\\

Multiple scattering of photons is important in the formation of spectral lines of interstellar media \cite{Dansac2020} and other astrophysical systems for which the escape of radiation after a  single large step is important \cite{Osterbrock1962,Adams1972}, a characteristic of a Lévy flight random walk \cite{Vezzani2019}. Recent works have proposed to use Lévy flight theories to interpret radiative transfer through resonant atomic vapors \cite{Lopez2023,Ivanov2024}. In \cite{Lopez2023}, for instance, expressions of transmittance as a function of thickness of a vapor and starting point of the random walk \cite{Buldyrev2001,Buldyrev20012,Klinger2022} were used to interpret transmission spectra through a vapor cell. The description of atomic spectra using Lévy flight theories depends on the details of the step-size distribution. We hope that the results presented in this article may help the development of the use of Lévy flight theory to interpret atomic spectra and radiative transfer, for instance, in astrophysical media and laboratory experiments.

\section*{Acknowledgments} 

M.O.A. thanks Universidade Federal de Pernambuco (UFPE) for financial support.I.C.N, J.P.L. and T.P.d.S. acknowledge financial support from Coordena\c c\~ao de Aperfei\c coamento de Pessoal de N\'ivel Superior (CAPES), Conselho Nacional de Desenvolvimento Cient\'ifico e Tecnol\'ogico (CNPq, Public Call
CNPq/MCTI/FNDCT Nº 18/2021)  and Para\'iba State Research Foundation (FAPESQ, grant n. 2021/3218). This work was funded by the Public Call n. 03 Produtividade em Pesquisa PROPESQ/PRPG/UFPB grant n. PVA13235-2020.



\end{document}